\documentclass[12pt,a4paper]{article}
\usepackage{amsmath}
\usepackage{amsthm}
\usepackage{epsfig}
\usepackage{latexsym}
\usepackage{pdflscape}
\numberwithin{equation}{section}
\usepackage{graphicx}
\title{Assessing the Inequalities of Wealth in Regions: \\ the
Italian Case}
\author{ 
  Roy Cerqueti$^{1,\#}$ and  Marcel Ausloos$^{2,3,*}$}
\date{ 
 $^1$ Department of Economics and Law, \\University of Macerata, \\Via Crescimbeni, 20, I - 62100  Macerata, Italy.\\ $^\#$email:  roy.cerqueti@unimc.it     \\  
   $^2$ eHumanities group, \\Royal Netherlands Academy of Arts and Sciences,
 Joan Muyskenweg 25, 1096 CJ Amsterdam, The Netherlands.\\ $^3$  Res. Beauvallon, rue de la Belle Jardini\`ere, 483/0021\\
B-4031, Li\`ege Angleur, Euroland \\$^*$email: marcel.ausloos@ulg.ac.be\\  }
\begin{document}

\maketitle

\begin{abstract}

This paper discusses  region wealth size distributions,   through
their member cities aggregated tax income. As an illustration, the
official data of the Italian Ministry of Economics and Finance has
been considered,  for all Italian municipalities, over the period
2007-2011.  Yearly data  of the aggregated tax income is transformed
into a few indicators: the  Gini, Theil, and Herfindahl-Hirschman
indices. On one hand, the relative interest of each index is
discussed. On the other hand, numerical results confirm that Italy
is  divided into  very different regional realities, a few  which
are   specifically outlined. This shows the interest of transforming
data in an adequate manner and of comparing such indices.

\end{abstract}

\section{ Introduction}\label{Introduction}
Spatial patterns based on geographical agglomerations and
dispersions of economic quantities play a fundamental role. In
discussing the features of the geographical entities, the
contribution that each city gives to the GDP of the reference
Country may be of particular interest.
\newline
The main purpose of the reported research here below  is to provide
a detailed analysis, both at a national as well as at the regional
level, of the  value (=size) wealth distribution among cities,
according to their Aggregated Tax Income, denoted hereafter ATI. The
numerical analysis is carried out on the basis of official data
provided by the Italian Ministry of Economics and Finance (MEF), and
concerns  each year of the 2007-2011 quinquennium. 
\newline
To pursue the scope, some statistically meaningful indicators are
computed. In particular, the Herfindahl index is calculated, while
adapted both Theil and Gini indices are provided.
\newline
The Theil index   (Theil 1967) represents one of the most common
statistical tools to measure inequality among data (Miskiewicz 2008,
Iglesias and de Almeida 2012, Clippe and Ausloos 2012). Basically
the index represents a number which synthesizes the degree of
dispersion of an agent  in a population with respect to a given
variable (= measure).
\newline The most relevant field of
application of the Theil index is represented by the measure of
income diversity. Therefore, it seems to be particularly appropriate
to compute such an  index here, ATI data being  a proxy of the
aggregated income of the citizens, clustered within cities or
regions, thereby representing municipalities  inhabitants wealth
diversity.
\newline
The Herfindahl index, also known as the Herfindahl-Hirschman index
(HHI), represents a measure of concentration (for some details on
the story of this index, developed independently by Hirschman in
1945 and Herfindahl in 1950, see Hirschman, 1964). It is applied
mainly to describe company  sizes  (in terms of $concentration$)
with respect to the entire market, and may then well represent the
amount of concentration  among firms (Alvarado 1999, Rotundo and
D'Arcangelis 2014). It is adapted here to the case of the ATI of
cities. Thus,  the HHI is an indicator of the amount of competition
among municipalities in a region, province,  or in the entire
country. The higher the value of HHI, the smaller the number of
cities with a large value of ATI, the weaker the competition in
concurring to the formation of Italian GDP. (From an industry
competition point of view, a HHI index below 0.01 indicates a highly
homogeneous index. From a portfolio point of view, a low HHI index
implies a very diversified portfolio).
\newline
The Gini index (Gini 1909) can be viewed as a measure of the level
of fairness of a resource distribution among a group of individuals
(Souma 2012, Bagatella-Flores et al. 2014, Aristondo et al. 2012).
\newline
To sum up, the Herfindahl index allows to gain insights on the level
of competition among cities and on their interactions, while both
Theil and Gini indices provide   measures of the dispersion of the
data. Since the analysis is performed not only at the country but
also at the regional level, such indices lead to a deeper
understanding of the Italian cities distribution at a global and
local level.
\newline
To the best of our knowledge, the analysis methods here employed
have not often been compared (but see Mussard et al. 2003), and
surely never  been applied to the Italian reality. Nevertheless, it
is fair to emphasize that  several contributions have appeared in
the literature for measuring the income inequalities in other
regional realities. In this respect, we mention Fan and Sun (2008)
for the measure of inequality in China over the period 1979-2006,
Walks (2013) for Canada, Bartels (2008) for the U.S.A., Wang et al.
(2007) for China. Some papers propose the statistical measure of the
income distribution in developing and poor Countries, which is an
interesting theme also for improving the economic growth of the
depressed areas (see e.g. Essama-Nssah, 1997 and Psacharopoulos et
al., 1995).
\newline
The paper is organized as follows: Section \ref{8092} contains the
description of the data. The adapted definition and computation of
the statistical indicators is found in Section \ref
{3economindices}. The findings are collected and discussed  in
Section \ref{sec:results}. The last section allows us to conclude
and to offer suggestions for further research lines. All the Tables
collecting the results at the regional level are reported in the
Appendix.
\section{Data}\label{8092}

The economic data analyzed  here below has been obtained by (and
from) the Research Center of the Italian MEF.  We have disaggregated
contributions at the  municipal level (in IT a \textit{municipality}
or \textit{city} is denoted as \textit{comune}, - plural $comuni$)
to the Italian GDP, for  five recent years: 2007-2011, in order to
keep the discussion as up-to-date as possible.
  \newline Let it be known that Italy (IT) is
composed of 20 regions, more than 100 provinces and 8000
municipalities.  Each municipality belongs to one and only one
province, and each province is contained in one and only one region.
Administrative modifications  due to the IT political system has led
to a varying number of provinces and municipalities during the
quinquennium, and also of the number of cities in each entity.  The
number of cities has been yearly evolving   as follows : 8101, 8094,
8094, 8092, 8092,  -  from 2007 till 2011.  In brief, several
(precisely 10) cities have  merged into  (3) new entities,  (2)
others were phagocytized.
\newline
First of all, it is worth to point out that 228  municipalities have
changed from a province to another one,  nevertheless remaining in
the same region, but 7 municipalities have changed from a province
to another one,   in  so doing also changing from a region (Marche)
to another one (Emilia-Romagna), in 2008.
\newline
However, the number of regions has been constantly equal to 20,
which makes the regional level  the most  interesting one for any
data measure and discussion.
\newline
Thus, we have considered the latest 2011 "count" as the basic one.
We have made   a virtual merging of cities, in the appropriate
(previous to 2011) years, according to IT administrative law
statements (see also $http://www.comuni-italiani.it/regioni.html$),
in order to compare ATI data for  "stable size" regions
 \newline In short,
the ATI of the resulting cities, thus {\it in fine}  for the regions
also,  have been linearly adapted, as if these were preexisting
before the merging or phagocytosis. Even this approximation is
reasonable for the negligible entity -in terms of regional ATI- of
the administrative changes, it seems to be of interest to further
investigate the economic effects of such modifications at a regional
level.
\newline Therefore,  the number of cities belonging
to a region   can  be summarized as in Table
\ref{TableNcityperregion}. For setting up the numerical analysis
framework, let   a summary of   the statistical characteristics for
ATI of  all IT cities ($N=8092$) in 2007-2011 be reported in Table
\ref{Tablestat} \footnote{The display of  the distribution
characteristics of these cities for the 110 provinces  would
obviously request  110  Tables (or Figures). They are not given
here, but any province case can be available from the authors, -
upon request.}.
\newline
Note that, in this time window, the data  claims a number of 103
provinces in 2007, with an increase by 7 units (institutionally
labeled as BT, CI, FM, MB, OG, OT, VS, which stand for
Barletta-Andria-Trani, Carbonia-Iglesias, Fermo, Monza e Brianza,
Ogliastra, Olbia-Tempio, Medio Campidano, respectively) thereafter,
leading to 110 provinces. In this respect, it is worth noting a
discrepancy between what data say and the real legislative evolution
of the provinces. In fact, 4 new provinces have been instituted by
the  12 July 2001 regional law in Sardinia and became operative in
2005 (CI, MB, OG, OT), while the 3 BT, FM and VS provinces have been
created on June 11, 2004 and became operative on June 2009. The
number of provinces was then changing : 103, 110, 110, 110, 110 -
from 2007 till 2011 for the statistical purpose of the MEF. In this
respect, it is interesting to observe that the Italian Government is
currently seeking for a reduction of the number of the provinces or,
eventually, their removal from the Italian Institutional setting.

\begin{table} \begin{center}
\begin{tabular}[t]{ccc}
  \hline  &$N_{c,r}$ \\ \hline
 Lombardia& 1544\\
Piemonte    &1206\\
Veneto  &581\\
Campania&   551\\
Calabria&   409\\
Sicilia&    390\\
Lazio   &378\\
Sardegna    &377\\
Emilia-Romagna&    348\\
Trentino-Alto Adige&    333\\
Abruzzo&    305\\
Toscana &287\\
Puglia& 258\\
Marche  &239\\
Liguria&    235\\
Friuli-Venezia Giulia&  218\\
Molise  &136\\
Basilicata  &131\\
Umbria& 92\\
Valle d'Aosta&  74\\  \hline
\end{tabular}
\caption{Number $N_{c,r}$ of (8092) cities (in 2011) taken into
account for calculating the economic indicators of   the (20) IT
regions in 2011, but in 2007-2010 as well, as explained in the
text. The regions are ranked according to the decreasing $N_{c,r}$ } \label{TableNcityperregion}
\end{center} \end{table}

\begin{table} \begin{center}
\begin{tabular}[t]{ccccccc}
  \hline
   $ $   &2007 &2008 &2009&2010&   2011 & \\
\hline
min. (x$10^{-5}$)   &3.0455         &2.9914      &   3.0909    &3.6083        &3.3479&   \\
Max. (x$10^{-10}$)&   4.3590&4.4360 &    4.4777      &4.5413 &4.5490 \\
Sum (x$10^{-11}$)&6.8947 &7.0427 &    7.0600 &7.1426 &7.2184      \\
mean ($\mu$) (x$10^{-7}$)   &8.5204  &8.7033     &   8.7248&8.8267 &8.9204  \\
median ($m$) (x$10^{-7}$)  &2.2875 &2.3553 &   2.3777 &2.4055&2.4601  \\
RMS (x$10^{-8}$)    &6.5629 &6.6598 &    6.6640&6.7531 &6.7701 \\
Std. Dev. ($\sigma$) (x$10^{-8}$) &6.5078&6.6031&   6.6070& 6.6956 &6.7115 \\
Var.    (x$10^{-17}$)&4.2351&4.3601&    4.3653 &4.4831 &4.5044 \\
Std. Err. (x$10^{-6}$)&7.2344 &7.3404  &   7.3448&7.4432 &7.4609 \\
Skewness  &48.685 &48.855&   49.266&49.414 &49.490 \\
Kurtosis      &2898.7     &2920.42   &   2978.1       &2991.0 &2994.7      \\  \hline
 $\mu/\sigma$  &0.1309   &0.1318&0.1321 &0.1319&  0.1329&  \\
$3(\mu-m)/\sigma$  &0.2873&0.2884&0.2883 &0.2878& 0.2889&   \\
\hline
\end{tabular}
   \caption{Summary of  (rounded) statistical
characteristics  for ATI (in  Euros) of IT cities ($N=8092$) in
2007-2011.}\label{Tablestat}
\end{center} \end{table}

 \begin{figure}
 \includegraphics[height=18cm,width=15cm]{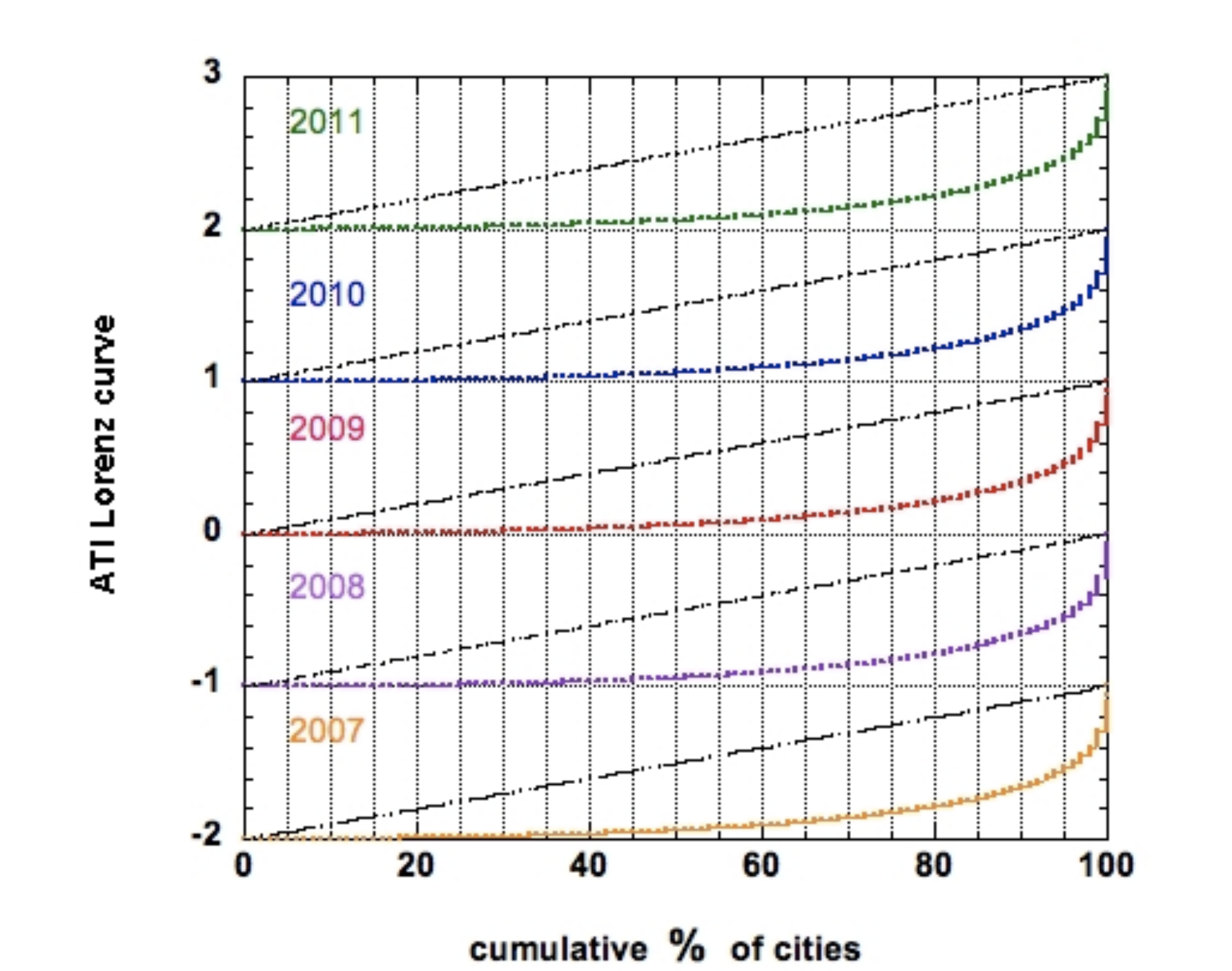}
 \caption{ ATI Lorenz curve  (color) and (straight) line of  ATI
 equality (black dots) for  various years of the  whole set of
 $N$=8092 cities, i.e. all in Italy in 2011.  The Gini coefficient (the values given in Table \ref{TableTheilHHGiniall})  is the area between those two lines. The
 Lorenz curves have been displaced along the $y$-axis by an obvious
 amount for  better readability.}   \label{fig:GinitotalIT}
 \end{figure}

\section{Statistical dissimilarity and competition among municipalities} \label{3economindices}

This section discusses whether cities exhibit on average similar ATI
values, and whenever their level of
 competition is high or low. For this purpose, the Theil, Gini
and Herfindahl indices for each IT region are computed. All the
results (necessarily numbers) are going to be presented in Tables
\ref{TableGiniAbruzzo}-\ref{TableGiniVeneto}.
Nevertheless, to exemplify what  a relevant corresponding figure,
for these indices would be,  see Fig.  \ref{fig:GinitotalIT} showing the Gini case for the whole Italy. Any reader should agree that it  seems unnecessary to present 20 similar figures.
Further discussion is postponed to Section \ref{sec:results}.
\newline
First, let us enter the various measure  details, with defining
formulae for completeness.

\subsection{Theil index}\label{theil} The Theil index  (Theil 1967)
is  adapted here as being defined by
\begin{equation}\label{Theilindexeq}
    Th =\frac{1}{N}\sum_{i=1}^N \frac{y_i}{\sum_j y_j} \cdot
\ln\left( \frac{y_i}{\sum_j y_j} \right)
\end{equation}
where $y_i$ is the  ATI  of the $i$-th city, and the sum $\sum_j
y_j$ is the aggregation of ATI in the entire Italy, while $N$ is the
number of cities ($N=8092$).
\newline It can be easily shown, from Eq. (\ref{Theilindexeq}),
that the  Theil index ($Th$)  is given by the difference between the
maximum possible entropy of the data and the observed entropy. It is
a special case of the generalized entropy index. The value of the
Theil index is then expressed in terms of negative entropy.
Therefore, a high Theil number indicates more order that is further
away from the "ideal" maximum disorder, $ln(N)$. More specifically,
a high level of Theil index is associated to a high distance from
the uniform distribution of the reference variable to the elements
of the sample set, which implies closeness to a polarized
distribution and a high level of dispersion. \newline Formally,
introducing the entropy:
\begin{equation}\label{entropyeq}
   H = -\sum_i^N \frac{y_i}{\sum_j y_j} \cdot \ln \left(\frac{y_i}{\sum_j y_j}\right)
\end{equation}
where $\frac{y_i}{\sum_j y_j}$ is the "market share" of the $i$-th
city, it  results that
\begin{equation}\label{HTheq}
   H \;=\;ln (N) - Th \;\;or\;\; Th= ln(N) - H.
\end{equation}

The order of magnitude of the Theil index is $Th\sim 1.73$ for the whole country; see Table \ref{TableTheilHHGiniall}.

\subsection{Herfindahl index}\label{Herfindahl}
The adapted  Herfindahl-Hirschman index (Hirschman 1964) is formally defined as follows:
\begin{equation}\label{HHIeq}
HHI =  \sum_{i\in L_{50}} \left(\frac{y_i}{\sum_j y_j} \right)^2,
\end{equation}
where $L_{50}$  is the set of the 50 largest cities in terms of ATI,
and $y_i$ is the ATI of the $i$-th city. The value 50 is
conventional, and  $HHI$  in Eq. (\ref{HHIeq}) is the sum of the
squares of the  market shares  of the 50 largest cities, where the
market shares are expressed as fractions. The index emphasizes the
weight of the largest cities. In our specific case, its value is
approximately given by $ 7\cdot 10^{-3}$;   see Table
\ref{TableTheilHHGiniall}. \newline A normalized Herfindahl index is
sometimes used and defined as:
\begin{equation}\label{NormHH}
H^{*} =  \frac{\left (HHI - 1/N \right )}{ 1-1/N }.
\end{equation}
with the appropriate $N$.  For $N$ large, $HHI  \simeq H^{*}$, as it is of course for the whole country. However, except for Lombardia and Piemonte, $N$ is usually less than 600. Thus, the "normalized" $H^{*}$ is also given in the economic index  tables,  for each region.

\begin{table} \begin{center}
  \begin{tabular}{|c|c|c|c|c|c|c|}
   \hline
whole IT   &   2007 & 2008&2009&2010& 2011&$<5yav>$  \\\hline
Entropy ($H$)&  7.2476  &   7.2603  &   7.2659  &   7.2669  &   7.2826  &   7.2650   \\
Max. Entropy &  8.9986  &   8.9986  &   8.9986  &   8.9986  &   8.9986  &   8.9986 \\
Theil index&    $\sim 1.751$   &1.7383    &1.7327    &1.7317 &1.7160
&1.7336  \\ \hline
10$^3$   $HHI$ &7.332    &   7.236   &   7.205   &   7.230   &   7.115   &   7.222  \\
10$^3$   $H^{*}$ &  7.209   &   7.113   &   7.083   &   7.107   &
6.992   &   7.099    \\ \hline
    Gini Coeff.&0.7591  &   0.7576 &    0.7566 &    0.7565 &    0.7547& 0.75685  \\
\hline
\end{tabular}
\caption{Statistical characteristics of the whole IT ATI data
distribution as a function of time; $N$= 8092. Entropy is $H$ (see
Eq. (\ref{HTheq})); Max. Entropy $\equiv ln(8092)$; the Herfindahl
index is $HHI$ (see Eq. (\ref{HHIeq})); the normalized Herfindahl
index is $H^{*}$ (see Eq. (\ref{NormHH})). Theil index is taken from
Eq. (\ref{Theilindexeq}). }\label{TableTheilHHGiniall}
 \end{center}
 \end{table}

\subsection{Gini coefficient}\label{gini}

Referring to the specific case treated here, the Gini index (Gini 1909)
can be defined through the so called  Lorenz curve, which  (in the present case) gives the proportion
$f$ of the total Italian ATI that is cumulatively provided by the
bottom $x$\% of the cities. If the Lorenz curve is a line at 45 degrees
in an $ f(x)$ plot, then  there is perfect equality of ATI. The Gini index (also called
coefficient  ($Gi$)) is the ratio of the area that lies between the line of
equality and the Lorenz curve over the total area under the line of
equality.
\newline A Gini coefficient of zero, of course,  expresses perfect
equality, i.e. all ATI values are the same, while a Gini coefficient
equal to one (or 100\%) expresses maximal inequality among values,
e.g.  only one city contributes to the the total Italian ATI.
\newline For example, the IT  Gini coefficient can be deduced from Fig.
\ref{fig:GinitotalIT} for the whole Italy. Its yearly value, given
in Table \ref{TableTheilHHGiniall},  is $\simeq 0.75$.  It is seen that  the IT  $Gi$  does not
much vary with time for the considered years.

\subsection{Local level coefficients}\label{localcoeff}
Each Theil, Gini and Herfindahl index can be calculated at the
country level, as presented in Table \ref{TableTheilHHGiniall} and also for each IT region  (or IT  province). All
formulae are easily transcribed. Nevertheless, for example, see how
the Gini coefficient for a region  reads:
\begin{equation}\label{ginicoeffreg}
Gi_{r}= \frac{2 \sum_{i=1}^{N_{c,r}} i\; y_{i,r}}{N_{c,r}\;
\sum_{i=1}^{N_{c,r}} y_{i,r}Ê} - \frac{N_{c,r} +1}{N_{c,r}}
\end{equation}
where $N_{c,r}$ is the number of cities in   region $r$ and
$y_{i,r}$ is the  ATI of the $i$-th city in region  $r$. Each $Gi_{r}$
is  given in the corresponding 20 Tables here below for each year.
\newline
The  Gini coefficient  for a province  would be
\begin{equation}\label{ginicoeffprov}
Gi_{p}= \frac{2 \sum_{i=1}^{N_{c,p}} i\; y_{i,p}}{N_{c,p}\;
\sum_{i=1}^{N_{c,p}} y_{i,p}} - \frac{N_{c,p} +1}{N_{c,p}}
\end{equation}
where $N_{c,p}$ is the number of cities in province $p$ and $y_{i,p}$ is
the  ATI of the $i$-th city in province $p$.
\newline
Similar writings hold for the Theil and Herfindahl indices.

\section{Results and discussion}\label{sec:results}

This section fixes and discusses the results of the investigation.
The 20 regional cases are reported in Tables
\ref{TableGiniAbruzzo}-\ref{TableGiniVeneto}. In exploring these
regional cases, several facts emerge. First, it
  is important to note a substantial time-invariance
of the values of the Theil, Herfindahl and Gini indices, which is
rather expected due to not excessive length of the examined time
interval.

The ranking of the Italian regions along the Theil, Gini
and Herfindahl indicator values leads to the identification of several
remarkable clusters. We discuss three of them.

\subsection{Region clustering through index values}
\begin{itemize}
\item Low Theil and Gini indices: Basilicata,
Valle d'Aosta, Puglia and Veneto are making the bottom four.
\end{itemize}
The   low levels of the Gini and Theil indicators  point to
regions with "fairly distributed" ATIs. Substantially, the member cities
contribute rather equally to their regional ATI.  Nevertheless, within this cluster, regional
cases can be quite diversified. In particular, Veneto is a region with a relevant
economic core, the so called \textit{Nord-Est}, with a great number
of rich mid-sized cities, -in terms of population,  which equally
share the regional economic market. In contrast, Valle d'Aosta is the smallest (in terms of $N_{c,r}$)
region of Italy: it  contains only 74 municipalities (with a small
number of inhabitants). A wide number of such cities are rich and
comparable in terms of ATI, and this explains the fairness of the
distribution of the ATI at a city level. However, Aosta -the main
city- is much larger than the other municipalities, and thereby  plays
a predominant role (look also at the high value of the HHI index for
Valle d'Aosta). Conversely, Puglia belongs to the South of Italy,
and its economic structure is still in development. For the
considered years, such a region appears to be made of small- and
mid-sized cities, - in terms of ATI. Hence, the fair distribution of
the ATIs among these municipalities describes here a generally low
level of individual city ATI values.

Therefore, there can be several "practical reasons" why an index is
small, and why a cluster can be somewhat of heterogeneous nature.

\begin{itemize}
\item Low Herfindahl index: Emilia Romagna, Marche, Puglia and Toscana exhibit the
lowest values.
\end{itemize}
This cluster mirrors different regional realities, which are however
comparable in terms of the economic competition among the cities.
Toscana has a large number of cultural and historical cities,
attracting an enormous flow of tourism. The industrial structure of
Toscana is also developed, and not polarized in a specific area. For
example, Prato (a small city close to Florence) is the headquarter
of a textile industrial district. Hence, the regional ATI is shared
among several not much populated cities. On the other hand, Emilia
Romagna has a peculiar economic characterization. The main part of
the business of this region is grounded on the food industry, which
is very delocalized in the entire territory. Amadori, one of the
largest companies in the sector of food in Italy, has  its
headquarter in San Vittore di Cesena, a very small municipality
close to Rimini. Yet, Bologna -the main city in Emilia Romagna- has
not  enough economic power to polarize the regional ATI of Emilia
Romagna.  Third, Puglia is a region whose economic structure is not
highly developed. In this case, the lack of competition is due to an
overall depressed situation.  Finally, Marche has plenty of
small-medium sized cities collecting extensive industrial districts.
The main economic activities of this region are also in this case
not concentrated in a small territory. They are principally based on
clothing and shoe factories. Several brands are worldwide famous,
like Diego Della Valle Tod's (the headquarter is located at Casette
d'Ete, a minuscule village close to Macerata) and Poltrona Frau
(headquarter in Tolentino, a little town in the center of Marche).
Ancona, the administrative center, is undoubtedly an important
harbor, but with more passengers than commercial activities. Hence,
Ancona is not economically powerful enough to polarize the regional
ATI of Marche.
\newline
Therefore,  the HHI index low value "cluster" also  implies
heterogeneity, but in a different manner than the Gini and Theil
index. Note that the only overlap between the two above clusters is
Puglia.

\begin{itemize}
\item High Theil, Gini and Herfindahl indices: Lazio and Liguria
assume the first two values of the rank, always, with very high values of
the indices. Piemonte belongs to this cluster for what concerns
Theil and Gini, but has a  HHI index rather small.
\end{itemize}
In this case, statistical indicators are coherent with the Italian
economical-geographical reality "common expectation". Lazio and
Liguria are polarized regions, where the main part of the ATI is
provided by a small number of cities. Specifically, there are two
municipalities (Rome for Lazio and Genova for Liguria), which are
remarkably predominant with respect to the other municipalities in
such regions. Is it worth recalling that Rome is the capital city of
Italy and  encloses also Vatican City (an independent State, but
with a huge percentage of Italians over the total labor force)? On
the other hand, Genova holds the main commercial harbor of Italy and
is the headquarter of very important companies and industrial units
(one for all: Finmeccanica SPA).
\newline The case of Piemonte is of great interest for
discussing indices through this cluster .
Piemonte exhibits high levels of Theil and Gini indicators, but HHI
is rather small. This fact meets the evidence that Turin, with the
FIAT company, provides the main part of the regional GDP. The low
Herfindahl index is due to the presence -among the high-rank fifty
municipalities- of a number of rich large-sized cities, but with rather
few inhabitants. Indeed, Piemonte has an important industrial
structure, and its economic market -in terms of ATI- is fairly
shared among several competitors.
\newline
Therefore, it is shown that there is some "practical value" in
discussing the three indices in parallel for a given region.

In concluding this subsection, note that the only overlap between
the two "low index" clusters is Puglia. In some sense, it could be
considered in itself as the extreme of the third cluster which have
values of the  indices.

\subsection{Evolutions}
The disorder in the yearly rankings of Italian regions for the
considered indicators is due to the oscillations of the
contributions that regions provide to the Italian GDP. However, the
rank changes are worth to be described.
\begin{itemize}
\item[$(i)$] For what concerns the Theil index, the
last two in 2007 (Puglia and Veneto) interchanged their position in
2011 ($r=19\rightarrow 20$, and conversely). In so doing, Veneto
lost its ever last place in the Theil index only, in 2011, but only
due to the 5th decimal;
\item[$(ii)$] in Theil index: Trentino Alto Adige ($r=10\rightarrow 11$) and Calabria  ($r=11\rightarrow
10$);
\item[$(iii)$] in Gini index:  Sardegna  ($r=4\rightarrow 5$) and Abruzzo  ($r=5\rightarrow
4$);
\item[$(iv)$] in  Gini index:  Sicilia  ($r=7\rightarrow 8$) and Umbria  ($r=8\rightarrow
7$);
\item[$(v)$] in Gini index, there is much reshuffling in row 13 to 16 between Calabria, Trentino Alto Adige, Friuli Venezia Giulia and
Emilia Romagna;
\item[$(vi)$] in HHI index:  Campania  ($r=8\rightarrow 9$) and Friuli Venezia Giulia  ($r=9\rightarrow
8$);
\item[$(vii)$] in  HHI index:   Trentino Alto Adige ($r=12\rightarrow 13$) and Lombardia  ($r=13\rightarrow
12$).
\end{itemize}
The changes in the rank listed above suggest to consider the
economic history of the considered regions, to grasp the reasons for
such modifications.  Two examples can illustrate the points:
\newline
\begin{itemize}
\item The case $(ii)$ can be explained by looking at the values of the
Theil indices in the corresponding Tables. Trentino Alto Adige moved
from 1.2823 (2007) to 1.2329 (2011), while Calabria from 1.2712
(2007) to 1.2344 (2011). The reduction of the Theil index means that
in both regions a more fair income distribution has been reached,
but Trentino Alto Adige was more unfair than Calabria in 2007. This
result can be interpreted as follows: the current financial crisis
has the merit of reducing the inequalities among the individuals,
even if such fairness is attained through an overall worsening of
the economic situation of Italy. The high economic level of Trentino
Alto Adige, which is richer than Calabria in terms of GDP {\it pro
capite}, is responsible of a more evident deterioration of the
economic situation at a regional level.
\end{itemize}
\begin{itemize}
\item
The change of position in $(vi)$ is due to a substantial invariance
of the Friuli Venezia Giulia's HHI (0.51994 in 2007, 0.51841 in
2011) and a remarkable decreasing in that of Campania (0.052596 in
2007 and 0.047086 in 2011). Campania is then over the quinquennium
increasingly less polarized, which suggests the tendency of the
cities to equally contribute to the regional ATI. This outcome is
due to the deterioration of the regional overall economic situation,
leading to the removal of the differences between the economic power
of the municipalities. In this respect, we recommend the reading of
the detailed report of the Bank of Italy regarding the economic
situation of Campania in 2011 (Bank of Italy 2011).
\end{itemize}

\subsection{Relative national impact}
Finally, it is interesting to point out how the indices values fare
with respect to the whole IT values. Note that
 \begin{itemize}
\item  for the Theil and Gini indices: Lazio, Liguria and Piemonte are
above the Italian  values of this indices, respectively;
\item for the HHI index: Lazio, Liguria,  but also Valle d'Aosta, Umbria and Molise
are above the Italian index value, but not Piemonte.
\end{itemize}
This result is expected for Lazio, Liguria and Piemonte (see the
discussion above). For what concerns Valle d'Aosta, Umbria and
Molise, the datum says that  a few cities highly polarize the
regional index (Aosta for Valle d'Aosta, Perugia and Terni for
Umbria and Campobasso and Isernia for Molise). The polarization is
due to different reasons: while Valle d'Aosta -a rich region-
collects a number of very small cities, leading to the predominance
of Aosta (which is by itself a rather small city, but much greater
than its competitors), Molise is a rather poor region where the
polarization is due to concentration of all the main institutions
(universities, companies' headquarters, political institutions) in
the most populated municipalities. Umbria is a particular case,
because polarization is due to the contribution of Perugia -the
capital of the region- but also to the presence of a very developed
industrial area -including also an important plant of the
Thyssenkrupp- close to Terni.

\section{Conclusions}\label{conclusions}
In this paper different classical economic indices have been adapted
and compared in order to emphasize their relative interest in
discussing city wealth contribution to a region wealth, - somewhat
as a function of  (recent) time. The analysis is supported through numerical
application as a statistical analysis of the Italian regions for the
period 2007-2011, measured by their municipalities aggregated tax
income values.

Thus, it has been shown, on the IT case, that it is of interest in
one hand to consider the (three) indices for a given region, and on
the other hand, to consider one index for a set of regions, and
compare the respective values. Moreover, it is of interest to
consider the relative values with respect to  the global set.

The data analysis confirms that IT is a unique entity, but with
different regional realities. In particular, a detailed description
of the 20 Italian regions through the Gini, Theil and
Herfindahl-Hirschmann indices contribute to explain the main
characteristics of the Northern and Southern regions. In particular,
we concur with Mussard et al. (2003) that the Gini index attributes
as much importance to the contribution between regions as to the
within-regions component, whereas the Theil and Herfindahl-Hirshmann
indices only consider that the inequalities are generated within the
regions.

\subsection*{Acknowledgements}
This paper is part of scientific activities in COST Action IS1104,
"The EU in the new complex geography of economic systems: models,
tools and policy evaluation".


\section*{Appendix}

This Appendix contains  the discussed economic indices of the 20 regional cases, in Tables
\ref{TableGiniAbruzzo}-\ref{TableGiniVeneto}.
\clearpage
 \begin{table} \begin{center}
 \begin{tabular}{|c|c|c|c|c|c|} \hline
Abruzzo  &   2007 & 2008&2009&2010& 2011  \\\hline
Entropy &   4.3501        &   4.3722      &   4.3586      &   4.3570      &   4.3634  \\
  Max. Entropy &   5.7203  &   5.7203      &   5.7203      &   5.7203      &   5.7203  \\
Theil index &   1.3702    &   1.3481     &   1.3617     &   1.3633    &   1.3569     \\
\hline
Herfindahl  &   0.033622        &   0.032324        &   0.032865        &   0.032937        &   0.032697    \\
  Norm. Herfindahl   &   0.030444        &   0.029141        &   0.029683        &   0.029756        &   0.029515    \\ \hline
Gini   Coeff.    &   0.750812        &   0.74967     &   0.75173     &   0.75212     &   0.75054 \\
 \hline
\end{tabular}
\caption{Various characteristics of the ATI data for Abruzzo as a
function of time; $N$= 305.}\label{TableGiniAbruzzo}
 \end{center}
 \end{table}

   \bigskip
\begin{table} \begin{center}
 \begin{tabular}{|c|c|c|c|c|c|} \hline
Aosta Valley &   2007 & 2008&2009&2010& 2011  \\ \hline
 Entropy    &   3.3003  &   3.3053  &   3.3129  &   3.3132  &   3.3204  \\
  Max. Entropy &   4.3041  &   4.3041  &   4.3041  &   4.3041  &   4.3041  \\
Theil index &   1.0037  &   0.99880 &   0.99114 &   0.99089 &
0.98369 \\ \hline
Herfindahl  &   0.10416 &   0.10330 &   0.10267 &   0.10244 &   0.10132 \\
  Norm. Herfindahl   &   0.091887    &   0.091017    &   0.090380    &   0.090150    &   0.089007    \\ \hline
Gini   Coeff.    &   0.64394 &   0.64290 &   0.63988 &   0.64020 &   0.63868 \\
\hline\end{tabular}
\caption{   Various characteristics of the ATI data for Aosta Valley
as a function of time; $N$=74.}\label{TableGiniAostaValley}
 \end{center}
 \end{table}

\clearpage

   \bigskip
\begin{table} \begin{center}
 \begin{tabular}{|c|c|c|c|c|c|}\hline
Basilicata &   2007 & 2008&2009&2010& 2011  \\\hline
Entropy &   3.8392  &   3.8544  &   3.8569  &   3.8493  &   3.8589  \\
  Max. Entropy &   4.8752  &   4.8752  &   4.8752  &   4.8752  &   4.8752  \\
Theil index &   1.0360  &   1.0208  &   1.0183  &   1.0259  & 1.0163
\\  \hline
Herfindahl  &   0.060062    &   0.058582    &   0.058448    &   0.059088    &   0.058246    \\
  Norm. Herfindahl   &   0.052831    &   0.051341    &   0.051205    &   0.051850    &   0.051002    \\  \hline
Gini   Coeff.    &   0.64826 &   0.64559 &   0.64498 &   0.64674 &   0.64461 \\
\hline\end{tabular}
\caption{Various characteristics of the ATI data for Basilicata as a
function of time; $N$= 131.}\label{TableGiniBasilicata}
 \end{center}
 \end{table}

\bigskip
  \begin{table} \begin{center}
 \begin{tabular}{|c|c|c|c|c|c|}\hline
Calabria&   2007 & 2008&2009&2010& 2011  \\\hline
 Entropy    &   4.7425  &   4.7614  &   4.7704  &   4.7729  &   4.7793  \\
  Max. Entropy &   6.0137  &   6.0137  &   6.0137  &   6.0137  &   6.0137  \\
Theil index &   1.2712  &   1.2523  &   1.2434  &   1.2408  & 1.2344
\\\hline
Herfindahl  &   0.031257    &   0.030534    &   0.030101    &   0.029947    &   0.029500    \\
  Norm. Herfindahl   &   0.028882    &   0.028158    &   0.027723    &   0.027569    &   0.027121    \\\hline
Gini   Coeff.    &   0.68512 &   0.68231 &   0.68072 &   0.68040 &   0.68055 \\
\hline
\end{tabular}
\caption{Various characteristics of the ATI data for  Calabria as a
function of time; $N$= 409.}\label{TableGiniCalabria}
 \end{center}
 \end{table}

    \bigskip
\begin{table} \begin{center}
 \begin{tabular}{|c|c|c|c|c|c|}\hline
Campania &   2007 & 2008&2009&2010& 2011  \\\hline
Entropy &   4.7335  &   4.7655  &   4.7739  &   4.7765  &   4.8062      \\
  Max. Entropy &   6.3117  &   6.3117  &   6.3117  &   6.3117  &   6.3117 \\
Theil Index &   1.5783  &   1.5463  &   1.5378  &   1.5352  & 1.5056
\\   \hline
Herfindahl  &   0.052596    &   0.049981    &   0.049289    &   0.049167    &   0.047086    \\
  Norm. Herfindahl   &   0.050873    &   0.048253    &   0.047561    &   0.047438    &   0.045354    \\  \hline
Gini   Coeff.    &   0.74246 &   0.73900 &   0.73812 &   0.73765 &   0.73390 \\
\hline\end{tabular}
\caption{Various characteristics of the ATI data for Campania as a
function of time; $N$= 551.}\label{TableGiniCampania}
 \end{center}
 \end{table}

             \bigskip
\begin{table} \begin{center}
 \begin{tabular}{|c|c|c|c|c|c|}\hline
Emilia Romagna &   2007 & 2008&2009&2010& 2011  \\\hline
Entropy &   4.6856  &   4.7011  &   4.7002  &   4.7032  &   4.7090      \\
  Max. Entropy &   5.8319  &   5.8522  &   5.8522  &   5.8522  &   5.8522  \\
Theil index &   1.1463  &   1.1511  &   1.1520  &   1.1490  & 1.1432
\\ \hline
Herfindahl  &   0.026274    &   0.025869    &   0.025875    &   0.025711    &   0.025425    \\
  Norm. Herfindahl   & 0.02341   &   0.023061    &   0.023068    &   0.022904    & 0.022617  \\ \hline
Gini   Coeff.    &   0.68118 &   0.68272 &   0.68254 &   0.68209 &   0.68066\\
\hline
\end{tabular}
\caption{Various characteristics of the ATI data for Emilia Romagna
as a function of time; $N$= 341 in 2007, and 348
next.}\label{TableGiniEmiliaRomagna}
 \end{center}
 \end{table}

 \bigskip
\begin{table} \begin{center}
   \begin{tabular}{|c|c|c|c|c|c|} \hline
Friuli Venetia Giulia &   2007 & 2008&2009&2010& 2011  \\\hline
Entropy &   4.1877      &   4.1849      &   4.179864        &   4.1799      &   4.1935\\
  Max. Entropy &   5.3891      &   5.3845      &   5.384495        &   5.3845      &   5.3845\\
Theil index &   1.2014     &   1.1996     &   1.2046     &
1.2046     &   1.1910    \\  \hline
Herfindahl  &   0.051994        &   0.052270        &   0.052972        &   0.052852        &   0.051841\\
  Norm. Herfindahl   &   0.047646        &   0.047902        &   0.048607        &   0.048487        &   0.047471\\ \hline
Gini   Coeff.    &   0.68181     &   0.68135     &   0.681851        &   0.68228     &   0.67983\\
 \hline
\end{tabular}
\caption{Various characteristics of the ATI data for Friuli Venetia
Giulia a function of time; $N$ depends on year. It is 219 in 2007,
and 218 next.}\label{TableGiniFriuli}
 \end{center}
 \end{table}

  \bigskip
\begin{table} \begin{center}
   \begin{tabular}{|c|c|c|c|c|c|} \hline
Lazio &   2007 & 2008&2009&2010& 2011  \\\hline
Entropy &   2.6121  &   2.6297  &   2.6458  &   2.6442  &   2.6664  \\
  Max. Entropy &   5.9349  &   5.9349  &   5.9349  &   5.9349  &   5.9349  \\
Theil index &   3.3228  &   3.3052  &   3.2891  &   3.2906  & 3.2685
\\  \hline
Herfindahl  &   0.37093 &   0.36688 &   0.36337 &   0.36350 &   0.35877 \\
  Norm. Herfindahl   &   0.36926 &   0.36520 &   0.36168 &   0.36181 &   0.35707 \\  \hline
Gini   Coeff.    &   0.88065 &   0.87985 &   0.87891 &   0.87927 &   0.87790 \\
\hline
\end{tabular}
\caption{Various characteristics of the ATI data for Lazio as a
function of time; $N$= 378.}\label{TableGiniLazio}
 \end{center}
 \end{table}

       \bigskip
\begin{table} \begin{center}
 \begin{tabular}{|c|c|c|c|c|c|} \hline
Liguria&   2007 & 2008&2009&2010& 2011  \\\hline
Entropy &   3.1712  &   3.1775  &   3.1859  &   3.1875  &   3.2039   \\
  Max. Entropy &   5.4596  &   5.4596  &   5.4596  &   5.4596  &   5.4596 \\
Theil index &   2.2884  &   2.2821  &   2.2737  &   2.2721  & 2.2557
\\ \hline
Herfindahl  &   0.19257 &   0.19169 &   0.19060 &   0.19010 &   0.18758 \\
  Norm. Herfindahl   &   0.18912 &   0.18824 &   0.18714 &   0.18664 &   0.18411 \\ \hline
Gini   Coeff.    &   0.83346 &   0.83257 &   0.83133 &   0.83143 &   0.82956 \\
\hline
\end{tabular}
\caption{Various  characteristics of the ATI data for Liguria as a
function of time; $N$= 235. }\label{TableGiniLiguria}
 \end{center}
 \end{table}

  \bigskip
\begin{table} \begin{center}
 \begin{tabular}{|c|c|c|c|c|c|}\hline
Lombardia &   2007 & 2008&2009&2010& 2011  \\ \hline
 Entropy    &   5.6933  &   5.7056  &   5.7140  &   5.7135  &   5.7239  \\
  Max. Entropy &   7.3434  &   7.3434  &   7.3434  &   7.3421  &   7.3421  \\
   Theil index &  1.6501  &   1.6379  &   1.6294  &   1.6287  &   1.6182  \\ \hline
Herfindahl  &   0.038857    &   0.038179    &   0.037639    &   0.037805    &   0.037402    \\
  Norm. Herfindahl   &   0.038235    &   0.037556    &   0.037016    &   0.037181    &   0.036779    \\ \hline
Gini   Coeff.    &   0.71799 &   0.71688 &   0.71592 &   0.71544 &   0.71405 \\
\hline
\end{tabular}
\caption{Various characteristics of the ATI data for Lombardia  as a
function of time; $N$=  1546 in 2007-09, and becomes 1544
next.}\label{TableGiniLombardia}
 \end{center}
 \end{table}

  \bigskip
\begin{table} \begin{center}
 \begin{tabular}{|c|c|c|c|c|c|}\hline
Marche &   2007 & 2008&2009&2010& 2011  \\ \hline
 Entropy    &   4.4416  &   4.4179  &   4.4165  &   4.4212  &   4.4328  \\
  Max. Entropy &   5.5053  &   5.4765  &   5.4765  &   5.4765  &   5.4765  \\
Theil index &   1.0638  &   1.0586  &   1.0600  &   1.0552  & 1.0436
\\ \hline
Herfindahl  &   0.024916    &   0.025284    &   0.025419    &   0.025215    &   0.024742    \\
  Norm. Herfindahl   &   0.020936    &   0.021189    &   0.021324    &   0.021119    &   0.020644    \\ \hline
Gini   Coeff.    &   0.70161 &   0.70162 &   0.70152 &   0.70082 &   0.69835 \\
\hline
\end{tabular}
\caption{Various characteristics of the ATI data for Marche as a
function of time; $N$= 239. }\label{TableGiniMarche}
 \end{center}
 \end{table}

        \bigskip
\begin{table} \begin{center}
 \begin{tabular}{|c|c|c|c|c|c|}\hline
Molise &   2007 & 2008&2009&2010& 2011  \\ \hline
 Entropy    &   3.6245  &   3.6314  &   3.6371  &   3.6407  &   3.6396  \\
  Max. Entropy &   4.9127  &   4.9127  &   4.9127  &   4.9127  &   4.9127  \\
Theil index &   1.2882  &   1.2813  &   1.2756  &   1.2719  & 1.2730
\\ \hline
Herfindahl  &   0.076722    &   0.076097    &   0.076336    &   0.076014    &   0.075998    \\
  Norm. Herfindahl   &   0.069883    &   0.069253    &   0.069494    &   0.069170    &   0.069153    \\ \hline
Gini   Coeff.    &   0.70074 &   0.69894 &   0.69593 &   0.69571 &   0.69669 \\
\hline
\end{tabular}
\caption{Various characteristics of the ATI data for Molise as a function of time; $N$=  136.}\label{TableGiniMolise}
 \end{center}
 \end{table}
  \bigskip
\begin{table} \begin{center}
 \begin{tabular}{|c|c|c|c|c|c|}\hline
Piemonte &   2007 & 2008&2009&2010& 2011  \\ \hline
Entropy &   5.0806  &   5.0873  &   5.0974  &   5.1005  &   5.1240  \\
  Max. Entropy &   7.0951  &   7.0951  &   7.0951  &   7.0951  &   7.0951  \\
Theil index &   2.0145  &   2.0077  &   1.9977  &   1.9946  & 1.9711
\\ \hline
Herfindahl  &   0.056106    &   0.055743    &   0.054883    &   0.054699    &   0.053053    \\
  Norm. Herfindahl   &   0.055323    &   0.054959    &   0.054099    &   0.053914    &   0.052267    \\ \hline
Gini   Coeff.    &   0.78607 &   0.78524 &   0.78395 &   0.78380 &   0.78179 \\
\hline
\end{tabular}
\caption{Various characteristics of the ATI data for Piemonte  as a
function of time; $N$=  1206. }\label{TableGiniPiemonte}
 \end{center}
 \end{table}

     \bigskip
\begin{table} \begin{center}
 \begin{tabular}{|c|c|c|c|c|c|}\hline
Puglia &   2007 & 2008&2009&2010& 2011  \\ \hline
 Entropy    &   4.5759  &   4.5907  &   4.5987  &   4.6018  &   4.6148  \\
  Max. Entropy &   5.5530  &   5.5530  &   5.5530  &   5.5530  &   5.5530  \\
Theil index &   0.9770  &   0.9622  &   0.9542  &   0.9512 &
0.9381  \\ \hline
Herfindahl  &   0.027864    &   0.027173    &   0.026807    &   0.026671    &   0.026025    \\
  Norm. Herfindahl   &   0.024082    &   0.023388    &   0.023020    &   0.022884    &   0.022235    \\ \hline
Gini   Coeff.    &   0.65401 &   0.65077 &   0.64895 &   0.64841 &   0.64556 \\
\hline
\end{tabular}
\caption{Various characteristics of the ATI data for Puglia  as a
function of time; $N$=  258. }\label{TableGiniPuglia}
 \end{center}
 \end{table}
    \bigskip
\begin{table} \begin{center}
 \begin{tabular}{|c|c|c|c|c|c|}\hline
Sardegna &   2007 & 2008&2009&2010& 2011  \\ \hline
 Entropy    &   4.4108  &   4.4302  &   4.4407  &   4.4405  &   4.4486  \\
  Max. Entropy &   5.9322  &   5.9322  &   5.9322  &   5.9322  &   5.9322  \\
Theil index &   1.5215  &   1.5020  &   1.4915  &   1.4918  & 1.4836
\\ \hline
Herfindahl  &   0.042023    &   0.040887    &   0.040300    &   0.040363    &   0.039878    \\
  Norm. Herfindahl   &   0.039475    &   0.038336    &   0.037747    &   0.037810    &   0.037325    \\ \hline
Gini   Coeff.    &   0.75202 &   0.74918 &   0.74738 &   0.74760 &   0.74684 \\
 \hline
\end{tabular}
\caption{Various characteristics of the ATI data for Sardegna as a
function of time; $N$=  377.}\label{TableGiniSardegna }
 \end{center}
 \end{table}
  \bigskip
\begin{table} \begin{center}
 \begin{tabular}{|c|c|c|c|c|c|} \hline
Sicilia &   2007 & 2008&2009&2010& 2011  \\ \hline
Entropy &   4.4933  &   4.5243  &   4.5327  &   4.5351  &   4.5584  \\
  Max. Entropy &   5.9661  &   5.9661  &   5.9661  &   5.9661  &   5.9661  \\
Theil index &   1.4729  &   1.4418  &   1.4335  &   1.4310  & 1.4077
\\  \hline
Herfindahl  &   0.045118    &   0.043540    &   0.043014    &   0.042918    &   0.041555    \\
  Norm. Herfindahl   &   0.042664    &   0.041081    &   0.040554    &   0.040457    &   0.039091    \\  \hline
Gini   Coeff.    &   0.73536 &   0.73044 &   0.72899 &   0.72878 &   0.72502 \\
\hline
\end{tabular}
\caption{ Various characteristics of the ATI data for Sicilia  as a
function of time; $N$=  390.}\label{TableGiniSicilia}
 \end{center}
 \end{table}

   \bigskip
\begin{table} \begin{center}
 \begin{tabular}{|c|c|c|c|c|c|} \hline
Toscana &   2007 & 2008&2009&2010& 2011  \\ \hline
 Entropy    &   4.5757  &   4.5809  &   4.5860  &   4.5875  &   4.5959  \\
  Max. Entropy &   5.6595  &   5.6595  &   5.6595  &   5.6595  &   5.6595  \\
Theil index &   1.0838  &   1.0786  &   1.0735  &   1.0719  & 1.0636
\\ \hline
Herfindahl  &   0.028356    &   0.028135    &   0.027904    &   0.027858    &   0.027605    \\
  Norm. Herfindahl   &   0.024959    &   0.024737    &   0.024505    &   0.024459    &   0.024205    \\ \hline
Gini   Coeff.    &   0.68612 &   0.68493 &   0.68342 &   0.68303 &   0.68126 \\
\hline
\end{tabular}
\caption{Various characteristics of the ATI data for Toscana as a
function of time; $N$=  287.}\label{TableGiniToscana}
 \end{center}
 \end{table}
  \bigskip
\begin{table} \begin{center}
 \begin{tabular}{|c|c|c|c|c|c|} \hline
Trentino Alto Adige &   2007 & 2008&2009&2010& 2011  \\\hline
Entropy &   4.5437  &   4.5386  &   4.5527  &   4.5619  &   4.5753\\
  Max. Entropy &   5.8260  &   5.8081  &   5.8081  &   5.8081  &   5.8081  \\
Theil Index &   1.2823  &   1.2695  &   1.2555  &   1.2462  &
1.2329\\    \hline
Herfindahl  &   0.039446    &   0.039042    &   0.038398    &   0.037800    &   0.036930    \\
  Norm. Herfindahl   &   0.036605    &   0.036147    &   0.035502    &   0.034902    &   0.034029    \\  \hline
Gini   Coeff.    &   0.68340 &   0.68286 &   0.68017 &   0.67910 &   0.67741 \\
\hline
\end{tabular}
\caption{Various characteristics of the ATI data for Trentino Alto
Adige as a function of time; $N$=333.
}\label{TableGiniTrentinoAltoAdige}
 \end{center}
 \end{table}

  \bigskip
\begin{table} \begin{center}
  \begin{tabular}{|c|c|c|c|c|c|} \hline
Umbria &   2007 & 2008&2009&2010& 2011  \\ \hline
 Entropy    &   3.3039  &   3.3074  &   3.3119  &   3.3117  &   3.3236  \\
  Max. Entropy &   4.5218  &   4.5218  &   4.5218  &   4.5218  &   4.5218  \\
Theil index &   1.2179  &   1.2144  &   1.2099  &   1.2101  & 1.1982
\\ \hline
Herfindahl  &   0.083292    &   0.082894    &   0.082372    &   0.082275    &   0.080978    \\
  Norm. Herfindahl   &   0.073219    &   0.072816    &   0.072288    &   0.072190    &   0.070879    \\ \hline
Gini   Coeff.    &   0.73213 &   0.73151 &   0.73072 &   0.73131 &   0.72907 \\
 \hline
\end{tabular}
\caption{Various characteristics of the ATI data for Umbria as a
function of time; $N$=  92.}\label{TableGiniUmbria}
 \end{center}
 \end{table}

    \bigskip
\begin{table}  \begin{center}  \begin{tabular}{|c|c|c|c|c|c|} \hline
Veneto &   2007 & 2008&2009&2010& 2011  \\ \hline
 Entropy    &   5.4028  &   5.4088  &   5.4073  &   5.4124  &   5.4266  \\
Max. Entropy &   6.3648  &   6.3648  &   6.3648  &   6.3648  &   6.3648  \\
Theil index &   0.9619  &   0.9559  &   0.9574  &   0.9524 &
0.9381  \\ \hline
Herfindahl  &   0.015352    &   0.015143    &   0.015164    &   0.015013    &   0.014591    \\
Norm. Herfindahl   &   0.013654    &   0.013445    &   0.013466    &
0.013314    &   0.012892    \\ \hline
Gini Coefficient    &   0.61816 &   0.61755 &   0.61821 &   0.61733 &   0.61476 \\
\hline
\end{tabular}
\caption{Various characteristics of the ATI data for Veneto as a
function of time; $N$=  581.}\label{TableGiniVeneto}
 \end{center}
 \end{table}

\clearpage

\begin{thebibliography}{99}

\bibitem{1}
Alvarado, F.L., 1999, Market Power: a dynamical definition,  Strategic Management Journal 20, 969-975

\bibitem{2}
Aristondo, O., García-Lapresta, J.L., Lasso de la Vega, C., Marques
Pereira, R.A., 2012. The Gini index, the dual decomposition of
aggregation functions, and the consistent measurement of inequality.
International Journal of Intelligent Systems 27(2), 132-152.

\bibitem{3}
Bagatella-Flores, N., Rodr\' iguez-Achach, M., Coronel-Brizio,H.F. ,
Hern\' andez-Montoya,   A.R. 2014, Wealth distribution of simple
exchange models coupled with extremal dynamics. (Unpublished
manuscript available at) $arXiv:1407.7153$.

\bibitem{4}
Bank of Italy, 2011, Economie regionali - L'economia della Campania.
Centro Stampa della Banca d'  IItalia.

\bibitem{5}
Bartels, L., Unequal Democracy: The Political Economy of the New Gilded Age. Princeton University Press: Princeton, 2008.

\bibitem{6}
Clippe, P., Ausloos, M., 2012. Benford's law and Theil transform of financial data, Physica A 391(24), 6556-6567.

\bibitem{ 7}
Fan, C.C., Sun, M., 2008. Regional Inequality in China, 1978-2006, Eurasian Geography and Economics 49(1), 1-20.

\bibitem{ 8}
Gini, C., 1909. Concentration and dependency ratios (in Italian).
English translation in Rivista di Politica Economica 87 (1997), 769-789.

\bibitem{ 9}
Hirschman, A.O., 1964. The paternity of an index, The American Economic Review 54(5), 761-762.

\bibitem{10 }
 Iglesias,  J. R.,    de Almeida, R.M. C.  2012,  Entropy and equilibrium state of free market models, European Journal of Physics B  85, 1-10.

\bibitem{ 11}
Miskiewicz, J., 2008. Globalization Entropy unification through the Theil index, Physica A 387(26), 6595-6604.

\bibitem{ 12}
Mussard, S., Seyte, Fr.,   Terraza, M. 2003. Decomposition of Gini and the generalized entropy inequality measures. Economics Bulletin,   4 (7) 1-6.

\bibitem{120}
Essama-Nssah, B., 1997. Impact of growth and distribution on poverty
in madagascar, Review of Income and Wealth 43(2), 239–252.

\bibitem{13}
Psacharopoulos, G., Morley, S., Fiszbein, A., Lee, H., Wood, W.C.,
1995. Poverty and income inequality in latin america during the
1980s, Review of Income and Wealth 41(3), 245–264.

\bibitem{14 }
Rotundo, G.,   D'Arcangelis, A.M.,  2014. Network of companies: an
analysis of market concentration in the Italian stock market,
Quality and Quantity     48 (4),  1893-1910.

\bibitem{ 140 }
Souma, R., 2002. Physics of Personal Income,  
in Empirical Science of Financial Fluctuations,  H. Takayasu ed, (Springer)
pp. 343--352

\bibitem{15 }
Theil, H., 1967. Economics and Information Theory, Chicago: Rand McNally and Company.

\bibitem{ 16}
Walks, A., 2013. Income Inequality and Polarization in Canada's Cities: An Examination and New Form of Measurement, Research Paper 227, Neighbourhood Change Research Partnership, University of Toronto, August 2013.

\bibitem{17}
Wan, G., Lu, M., Chen, Z., 2007. Globalization and regional income
inequality: empirical evidence from within china, Review of Income
and Wealth 53(1), 35–59.



\end{thebibliography}
\end{document}